\documentclass[preprint2]{aastex}
\usepackage{amsmath,amssymb}
\usepackage{graphicx}
\usepackage{rotating}
\usepackage{longtable}
\usepackage{hyperref}

\defcitealias{MTL1}{Paper~I}

\newcommand{\myemail}{kobelski@solar.physics.montana.edu}

\shorttitle{Forward Modeling Hi-C}
\shortauthors{Kobelski et al.}

\begin{document}

\title{Forward Modeling Transient Brightenings and Microflares around an Active Region Observed with Hi-C}

\author{Adam R. Kobelski \altaffilmark{1} and David E. McKenzie}
\affil{Department of Physics, PO Box 173840, Montana State University,
	Bozeman, MT, 59717-3840, USA}
	\email{\myemail}
\altaffiltext{1}{Now at National Radio Astronomy Observatory, P.O. Box 2, Green Bank, WV 24944}

\begin{abstract}
Small scale flare-like brightenings around active regions are among the smallest and most fundamental of energetic transient events in the corona, providing a testbed for models of heating and active region dynamics. In a previous study, we modeled a large collection of these microflares observed with {\it Hinode}/XRT using EBTEL and found that they required multiple heating events, but could not distinguish between multiple heating events on a single strand, or multiple strands each experiencing a single heating event. We present here a similar study, but with EUV data of Active Region 11520 from the High Resolution Coronal Imager (Hi-C) sounding rocket. Hi-C provides an order of magnitude improvement to the spatial resolution of XRT, and a cooler temperature sensitivity, which combine to provide significant improvements to our ability to detect and model microflare activity around active regions. We have found that at the spatial resolution of Hi-C ($\approx$0.3''), the events occur much more frequently than expected (57 events detected, only 1 or 2 expected), and are most likely made from strands of order 100 km wide, each of which is impulsively heated with multiple heating events. These findings tend to support bursty reconnection as the cause of the energy release responsible for the brightenings.

\end{abstract}


\section{Introduction}
The heating of active regions can be considered in two forms: the large impulsive heating seen in solar flares \citep[][]{Benz2008,ShibataMagara2011}, and a more constant background heating, such as from frequent nanoflaring \citep[][]{Parker1988} or spicules \citep[][]{DePontieuetal2007}. While it is fairly standard to associate flaring events to magnetic reconnection \citep[{\it e.g.}:][]{ForbesActon1996,Shibata1999,Fletcheretal2001,Yokoyamaetal2001,Qiuetal2010}, the source of the background heating is more hotly debated. One dominant discussion is the existence of small ``nanoflares'' as the source of this heating \citep[][among others]{Parker1988,Klimchuk2006} which act as flares on much smaller (spatial, temporal, and energetic) scales.   

One of the most sought after signatures of nanoflare heating is the existence of high temperature ($>$10 MK) plasma in quiescent active regions. There has been evidence for this high temperature component in data from the Ramaty High Energy Solar Spectroscopic Imager \citep[RHESSI,][]{McTiernan2009},  the {\it Hinode}/X-Ray Telescope \citep[XRT,][]{Realeetal2009,Schmelzetal2009a}, and Atmospheric Imaging Assembly (AIA) onboard the Solar Dynamics Observatory \citep[SDO,][]{ViallKlimchuk2011}, often using proxies of the differential emission measure (DEM) to estimate the spectral components of the observed plasma. However, this hot component as applied to nanoflare heating has been difficult to find with spectral observations from {\it Hinode}/Extreme Ultraviolet Imaging Spectrometer \citep[EIS,][]{Warrenetal2012} or High Resolution Coronal Imager \citep[Hi-C, ][]{Winebargeretal2013}.

While nanoflare heating may not be the sole source of heating in active regions, small flaring events (sometimes referred to as microflares) are quite readily visible \citep[][]{Linetal1984,Garyetal1997}. Below the GOES flare detection and labeling threshold, the frequency of coronal flare-like brightenings around active regions is quite large \citep[as many as 40 events per hour per active region,][]{BerghmansMcKenzieClette,Shimizuetal1992}. The rate of occurrence of these brightenings tends to increase as the energies involved decrease (down to the noise floor of the instrument used), though not necessarily at a rate sufficient to heat the corona by nanoflare heating \citep[][]{Hudson1991,Berghmans2002}. These active region transient brightenings (ARTBs) can be observed with Extreme Ultraviolet \citep[EUV,][]{Seaton2001} and soft X-ray instruments \citep[][]{Shimizuetal1992,Shimizu1995}, though their transient nature and the differences in temperature response make direct comparison between the two wavelengths difficult \citep[see for example:][]{BerghmansMcKenzieClette}. In this paper, we do not distinguish between ARTBs and microflares, and use the terms interchangeably. 

In order to study the scales involved in heating small scale coronal brightenings, \citet{MTL1} (hereinafter \citetalias{MTL1}),  detected and modeled 34 ARTBs observed with XRT. The ARTBs were modeled as bundles of independent strands using a variety of heating functions and the Enthalpy Based Thermal Evolution of Loops \citep[EBTEL, ][]{EBTEL1, EBTEL2, EBTEL3} code, incorporating a genetic algorithm to find the parameters that best reproduced the observed light curves of the brightenings. The results of \citetalias{MTL1} showed that an impulsive mechanism was most likely required to instigate the heating of the individual strands, though they could not distinguish between ARTBs heated by a single strand experiencing multiple heating events, or many strands each being heated once. 

The improved resolution of Hi-C provides a unique opportunity to better constrain the size of the events which cause these brightenings. From the observations of Hi-C the brightening loops appear to consist of multiple strands ``braided'' together, each of which are heated multiple times \citep[][]{CirtainHiC2013}. In this article, we apply the detection and modeling methodology of \citetalias{MTL1} to Hi-C data. By detecting and modeling ARTBs observed with Hi-C we have found that the rate of ARTBs might be significantly higher than previously discussed in the literature, consistent with the findings of \citet{Testaetal2013}. The results of this forward-model also support previous findings \citep[][]{CirtainHiC2013, Winebargeretal2013, Brooksetal2013} that the scale of the observed braiding is very close to or smaller than the resolution of the Hi-C instrument, promoting the need for more instruments with similar observational capabilities as Hi-C if we hope to understand the scales of energy release in the corona.

A brief overview of the method used to detect and model ARTBs is given in Sections~\ref{hic_detect} and \ref{hic_model}. Section~\ref{hic_data} discusses the observations, with the results of the study shown in Section~\ref{hic_results}. A discussion of these results and comparisons to other studies is given in Section~\ref{hic_discussion}. Section~\ref{hic_conc} provides a summary of conclusions.

\section{Method Overview}\label{hic_method}
The method of detection and modeling is thoroughly discussed in \citetalias{MTL1}, and we present here a brief overview for the benefit of the reader.
\subsection{ARTB Detection}\label{hic_detect}
To find ARTBs in sequences of coronal images, we use a slightly modified version of the detection scheme utilized by \citet{BerghmansClette}. The detection starts by subtracting the temporal running mean (width $w_{\rm rm}$) from a calibrated, exposure normalized, and aligned stack of images on a pixel by pixel basis. After dividing the running mean subtracted image by the standard deviation, the algorithm then looks for the brightest pixel in this residual. If this pixel value is larger than $q_{\rm D}$, neighboring pixels that are larger than $q_{\rm C}$ are grouped together. If 10 or more such pixels can be connected, a light curve is created, bounded by the largest spatial extent of the detected region. The flux of the light curve is normalized to DN s$^{-1}$ pixel$^{-1}$. Regardless of whether or not the detected region is larger than our 10 pixel threshold, the pixel's values within the detected region are reset to the median value, and the process repeated until the brightest pixel in the residual is less than $q_{\rm D}$. The extracted light curves are then modeled as multi- or single-stranded loops.

In \citetalias{MTL1}, we used $w_{\rm rm}=20$, $q_{\rm D}=4$, and $q_{\rm C}=3$ to detect ARTBs with the {\it Hinode} X-Ray Telescope {\citep[XRT:][]{Kano2008, Narukage2011, GolubXRT}. Since Hi-C data are of significantly higher resolution (temporally, spatially, and spectrally), we found that these detection parameters should be modified to efficiently detect ARTBs with Hi-C. Exploratory analysis runs led to satisfactory results with $w_{\rm rm}=15$, $q_{\rm D}=3.5$, and $q_{\rm C}=3$. The values of these detection constants caused significantly more spurious detections than in \citetalias{MTL1}, but due to the small size of the Hi-C data set, it was deemed easier to remove these detections manually (see Section~\ref{hic_data}). An additional difference between the detections in this study and those in our previous work is the detection (and analysis) of brightenings further from the core of the active region. These detections outside of the AR core generally occur along loops which appear anchored on one side to the AR, but could potentially behave slightly different than the ARTBs traditionally studied. For this reason, we also use the more general term, microflare.

\subsection{Multi-Stranded Model}\label{hic_model}
Using the 0 dimensional EBTEL framework, we then forward model the light curves of the detections as multi- or single-stranded loops with each strand modeled independently. The fluxes of the resultant strands are then superimposed to create a modeled loop. We then utilize the genetic algorithm pikaia  \citep[][]{pikaia} to traverse the parameter space of the model (see Table~\ref{param_spc}) to find which parameters minimize the $\chi^2$ value between the modeled ARTB flux and the observed ARTB flux from Hi-C. 

EBTEL requires a strand half-length ($l_{\rm s}$) and a heating function to model each strand. We estimate the (full) strand length ($L_{\rm obs}$) by measuring the projected length of the strand as observed with Hi-C, and then constrain pikaia to search for strand lengths $L_{\rm obs}/2.5\le l_{\rm s} \le 2.5L_{\rm obs}$: the upper limit accommodates a large range of projection effects, the lower limit allows a sanity check to the results. For each realization of the forward model, the strand length is fixed. The heating function for each individual strand is a triangular pulse whose width is varied by pikaia (though fixed for each loop realization) and whose peak is dictated by a heating envelope for the group. As in \citetalias{MTL1}, for the case of a multi-stranded loop, we test the capabilities of two heating envelopes, a sinusoidal and a ``lambda'' envelope. Figure~\ref{heatenv} (modified from \citetalias{MTL1}) illustrates these heating envelopes. The sinusoidal envelope shows a symmetric increase and decline for the rate of heat input to individual strands, such as might be predicted if the strands are heated by resonant absorption \citep[][]{Ofmanetal1995,WalshIreland2003}, and the lambda envelope suggests a more impulsive onset to strand heating (followed by a period of relaxation) such as might be expected from a bursty reconnection scenario. The delay between individual heating events is constant for a single loop realization, but is varied for different realizations by pikaia. We also model strands as monolithic loops, which act identical to the multi-stranded lambda envelope loops, except the heating all occurs within a single strand. Assuming a cylindrical strand of constant radius, the EBTEL results are then convolved with the instrument response function of Hi-C yielding a predicted flux. The strand radius of the model flux is then varied so that the peak flux from the model matches the peak flux of the observation. The multi-stranded sinusoidal and lambda envelopes, as well as the monolithic strand heated with the lambda envelope, are all independently tested, such that the results from each model can be compared. 

\begin{table*}\centering
\caption{Parameter space searched by pikaia. $L_{\rm obs}$ is the projected full length of the strand, as observed in the Hi-C images. The range of strand radii was not limited.}
\begin{tabular}{c|c|c}
\hline\hline
Min & Parameter & Max\\
\hline
0.05 & Peak Heating Rate (ergs cm$^{-3}$ s$^{-1}$) & 0.55\\
2 & Heat Pulse Width (s) & 27\\
0.4 & Strand Half-Length ($L_{\rm obs}$) & 2.5\\
1 & Event Delay (s) & 10\\
1 & Number of Heatings & 30\\
\hline
\end{tabular}\label{param_spc}
\end{table*}

\begin{figure*}[ht]
\includegraphics[width=1.\linewidth]{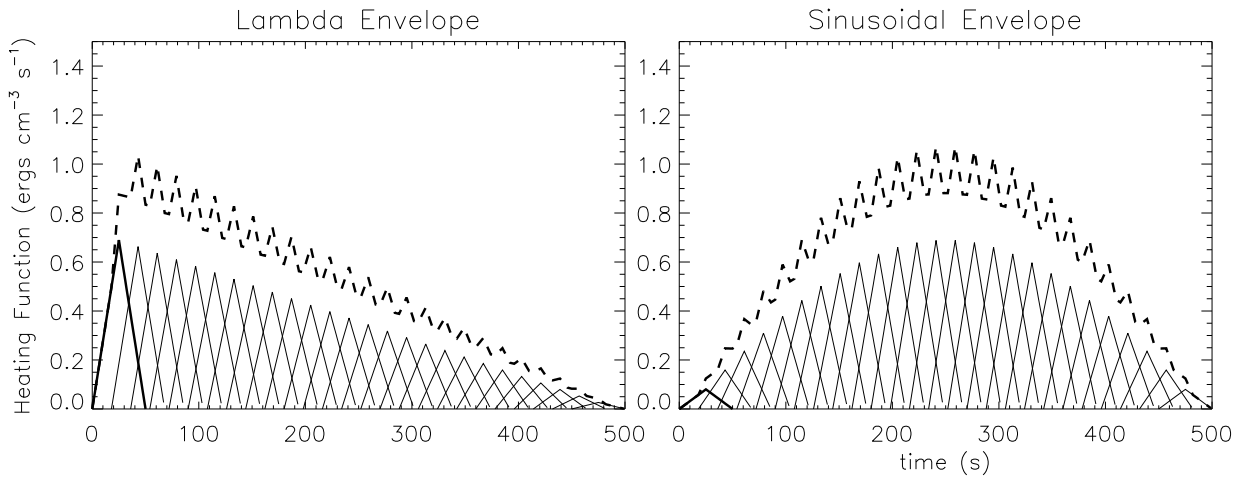}
\caption{An example of the heating function for the case of 26 individual strands with a heating delay of 12 s, heating width of 50 s and a peak heating rate of 0.69 ergs cm$^{-3}$ s$^{-1}$. On the left is the lambda shaped envelope, the sinusoidal envelope on the right. The solid lines represent the individual heating events, and the dashed line is the sum of the individual events such as would be used to heat the monolithic strand. These parameters (number of strands/heating events, $\Delta t$, heating width, and peak heating rate) are varied for each realization of the model by the genetic algorithm in order to find the best fit between the combined EUV flux of the strands and flux observed with Hi-C. This figure is a modified version of a similar figure shown in \citetalias{MTL1}.}
\label{heatenv}
\end{figure*}

\section{Data}\label{hic_data}
The Hi-C sounding rocket \citep[][]{CirtainHiC2013} was launched on 2012 July 11, taking data of Active Region 11520 for over 5 minutes, from 18:52:09 to 18:57:26 UT. An HMI magnetogram from near the time of the Hi-C launch is shown in Figure~\ref{hic_hmi}, with the approximate FOV of Hi-C denoted by the black box. The telescope is of very high resolution, taking images with 0.1'' pixels in a narrow wavelength bandpass around 193\AA. The entire 4096 $\times$ 4096 pixel CCD (4K) was read out from 18:52:09 to 18:55:30 at an average cadence of 5.7s image$^{-1}$, at which point the field of view was reduced to the central 1024 $\times$ 1024 pixels (1K) and the cadence increased to 1.38 s image$^{-1}$. The field of view was centered approximately [-130, -453] arcsecs from disk center. These data were then calibrated, including dark subtraction, flat field removal, dust spot removal, and co-alignment (including tracking). In the case of the higher cadence 1K data, 4 images were stacked to improve the signal to noise ratio, and to match the cadence of the larger images. 

\begin{figure}[ht]
\includegraphics[width=1.1\linewidth, clip=true, trim=20mm 0mm 20mm 0mm]{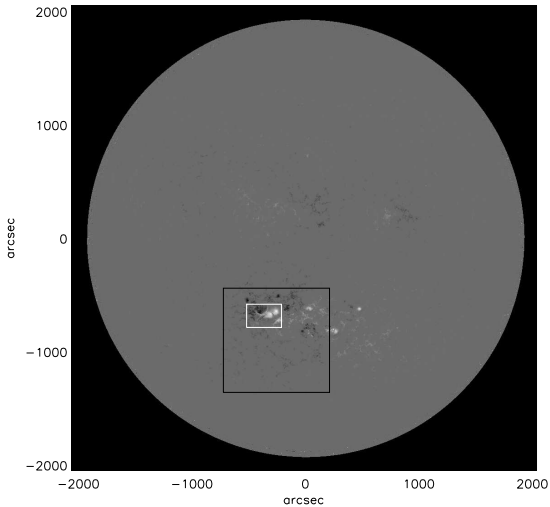}
\caption{Full Disk (720s) HMI magnetogram from 19UT on 2010 July 11. The approximate full Hi-C field of view is denoted by the black box, and the white box shows the smaller subregion shown in Figure~\ref{hic_detect_map}. The detection was run on the full Hi-C field of view.}
\label{hic_hmi}
\end{figure}

The detection algorithm was run on both the full field of view 4K Hi-C data as well as the higher cadence, smaller field of view 1K data. To ease the computational requirements, the 4K data were split into sixteen 1K data sets. As mentioned in Section~\ref{hic_detect}, the parameters used in the detection algorithm were not fully optimized, returning many spurious ``false positive'' detections. These spurious detections include spikes due to instrument noise, and flux enhancements existing where only the beginning or end of the brightening was observed. Further optimization of the parameters would have taken more time than manual removal of the spurious detections, so we opted to tune the algorithm to over-detect regions and remove the poor detections manually by the visual inspection of the light curve. In all, the algorithm returned 452 detections, 395 of which were spurious, and 57 were deemed suitable for analysis. The accepted detections are found in Table~\ref{hic_artb_tab}. Figure~\ref{hic_detect_map} shows a subframe of a Hi-C image and the detections within that region have been overplotted. The detected regions were then normalized as discussed in Section~\ref{hic_detect}, and background subtracted by removing 80\% of the lowest flux observed in the light curve (as discussed in \citetalias{MTL1}). The aspect ratios denoted in Table~\ref{hic_detect_map} were calculated as the square root of the eigenvalues of the second moment of inertia matrix of the detection mask. Low values of the aspect ratio (near unity) potentially suggest a footpoint source, while larger values (much greater than unity) illustrate a detection of a complete loop. This value, though, should be considered a minimum value, since it only represents the shape detected by the algorithm, and not necessarily the complete shape of the brightening. The range in these values suggest that we are seeing some footpoint heating events, and many full loop heatings.

\begin{table*}[]\scriptsize\centering
\caption{Information on ARTBs detected and analyzed from the Hi-C data set running from 18:52UT to 18:56UT on 2012 July 11. The large variation in ARTB sizes, with a few large and many smaller detections, helps to illustrate the advantage of using a high resolution instrument such as Hi-C. The aspect ratio was calculated from the shape of the detection, showing the range of shapes of detected regions, where values near one represent circular detections.}
\begin{tabular}{c|c|c|c||c|c|c|c}
\hline\hline
 ARTB & Size of ARTB & Number of & Aspect & ARTB & Size of ARTB & Number of & Aspect \\ 
 Number & (pixels/image) & Images & Ratio & Number & (pixels/image) & Images & Ratio \\
\hline
01 &  104 & 13 &   6.0&30 &   52 & 14 &   4.2\\
02 &   62 & 19 &   3.6&31 &   11 & 13 &   2.1\\
03 &   26 & 28 &   3.3&32 &   46 &  9 &   2.3\\
04 &   70 & 24 &   1.6&33 &   61 & 14 &   2.6\\
05 &   11 & 26 &   3.3&34 &    9 & 16 &   1.0\\
06 &   67 & 35 &   4.2&35 &   16 & 11 &   2.0\\
07 &   11 & 12 &   1.7&36 &   44 & 15 &   3.0\\
08 &   39 & 25 &   3.6&37 &   33 & 10 &   1.5\\
09 &   58 & 33 &   8.1&38 &   49 & 24 &   2.4\\
10 &   54 & 27 &  56.6&39 &   18 & 26 &   3.5\\
11 &   76 & 13 &   4.2&40 &    6 & 30 &   2.7\\
12 &   39 & 26 &   7.4&41 &    9 & 11 &   5.8\\
13 &   24 & 13 &  48.5&42 &   22 & 21 &   5.5\\
14 &   51 & 14 &   1.6&43 &   12 & 20 &  12.0\\
15 &   18 & 24 &   3.5&44 &   11 & 26 &   3.7\\
16 &   23 & 26 &   3.0&45 &   59 & 24 &   1.6\\
17 & 1072 & 13 &  29.7&46 &   22 & 26 &  28.0\\
18 &   46 & 16 &  14.5&47 &   17 & 21 &   4.0\\
19 &   40 & 17 &  15.2&48 &   19 & 23 &   6.9\\
20 &   84 &  6 &   7.7&49 &    9 & 21 &   2.7\\
21 &   70 & 11 &   9.7&50 &   14 & 23 &   5.5\\
22 &   16 & 19 &   3.0&51 &   26 & 32 &   2.5\\
23 &  520 & 14 &  10.9&52 &   11 & 25 &   3.7\\
24 &   58 & 10 &   3.4&53 &   12 & 21 &   2.9\\
25 &   95 &  9 &  45.8&54 &   34 & 19 &   3.1\\
26 &   12 & 21 &   2.9&55 &   13 & 10 &   4.2\\
27 &   21 & 18 &   5.4&56 &   22 & 26 &   4.6\\
28 &   86 & 18 &  11.6&57 &   24 & 20 &  16.9\\
29 &   17 & 13 &   1.2&\\
\hline
\end{tabular}
\label{hic_artb_tab}
\end{table*}

\begin{figure*}[t]
\includegraphics[width=1.\linewidth]{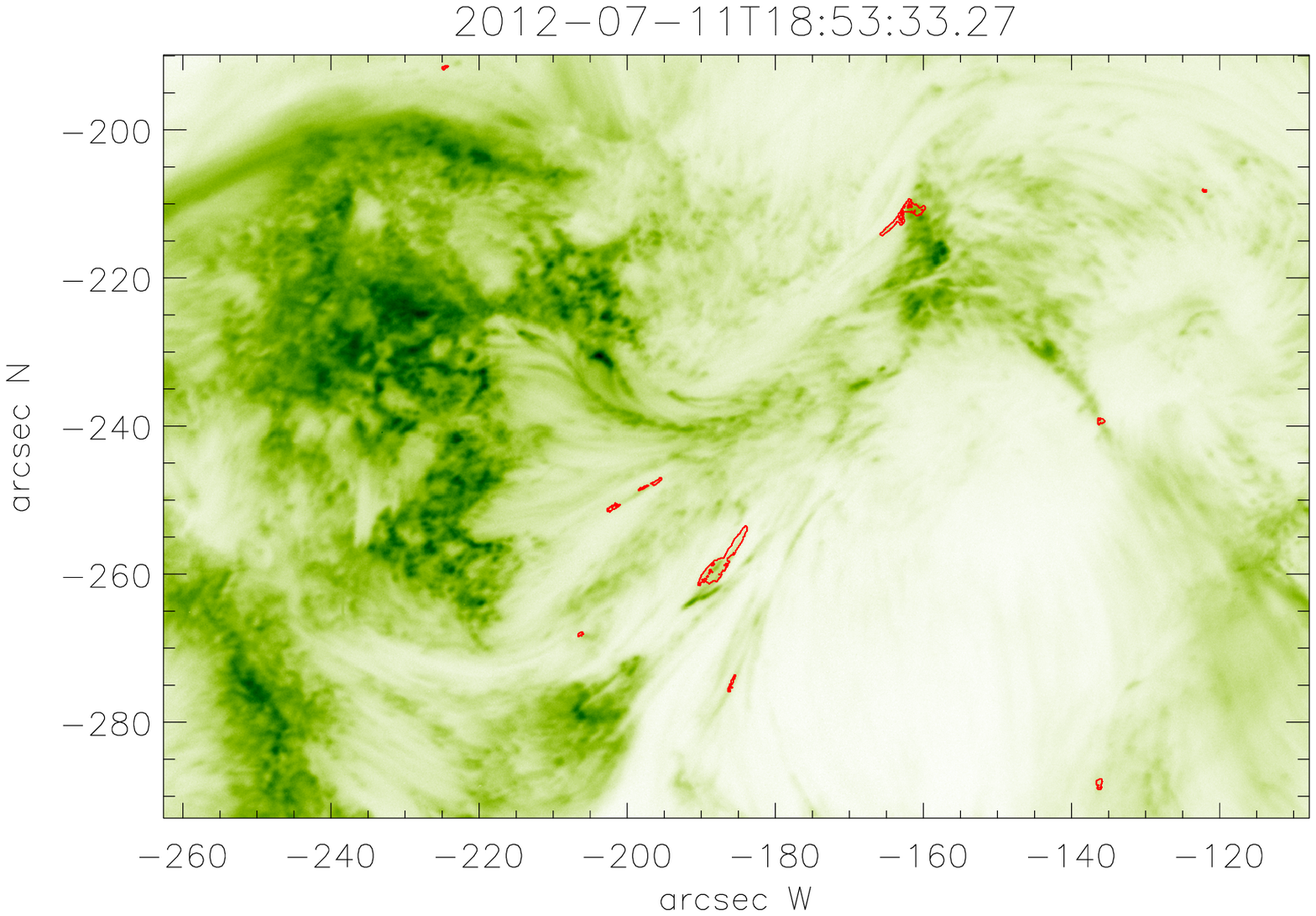}
\caption{Reverse color example sub-image (see Figure~\ref{hic_hmi}) from a full-frame Hi-C image, with contours of detected and analyzed ARTBs overlayed in red. Here you can see a spread of the size and shape of detections as noted in Table~\ref{hic_artb_tab}. The axis labels denote the distance from disk center in arcseconds. As shown in Figure~\ref{hic_hmi}, the left side is the trailing (diffuse) negative flux, and the right side is the leading positive flux.}
\label{hic_detect_map}
\end{figure*}

\section{Results and Analysis}\label{hic_results}
The 57 ARTBs detected were forward modeled as discussed in Section~\ref{hic_model}; typical fits are shown in Figures~\ref{hic_fit_examp} and \ref{hic_temps}. The parameters used to create these example fits are shown in Table~\ref{hic_exampfits_params}. The median results of all of the fittings for each envelope are shown in Tables~\ref{hic_rslts_tab} and \ref{hic_rslts_temp_tab}. The loop lengths noted in Table~\ref{hic_rslts_tab} represent the strands' length in the corona as used in the EBTEL calculation. The aspect ratio listed in Table~\ref{hic_rslts_tab} represents the model's resultant strand half-length divided by the strand radius. The multi-stranded model is allowed to create strands with radii the same size as their length, in essence creating point like brightenings, hence we use this aspect ratio as a validity test of our results, since it shows relatively slender strands.

The integral ratio shown in Table~\ref{hic_rslts_temp_tab} is defined as the ratio between the fluxes of the model and the observation integrated over the length of the observation. The $\chi^2$ value used is not normalized (hence the units of DN), and thus its magnitude is relative to the intensity of the observed microflare. As the $\chi^2$ should not be directly compared between different data sets, the integral ratio is an effective method for comparing the quality of fits between different events. The temperatures used in Table~\ref{hic_rslts_temp_tab} are the emission measure weighted average temperature, from which the duration of elevated temperatures are calculated. The duration of elevated temperatures is shown as the time spent above half-peak temperature (essentially the full-width at half-maximum, and denoted as such). Also shown in Table~\ref{hic_rslts_temp_tab} is the amount of time the emission measure weighted average temperature spends above 5 MK, i.e. how long the loop appears `hot.'

\begin{figure*}[ht]
\includegraphics[width=0.5\linewidth]{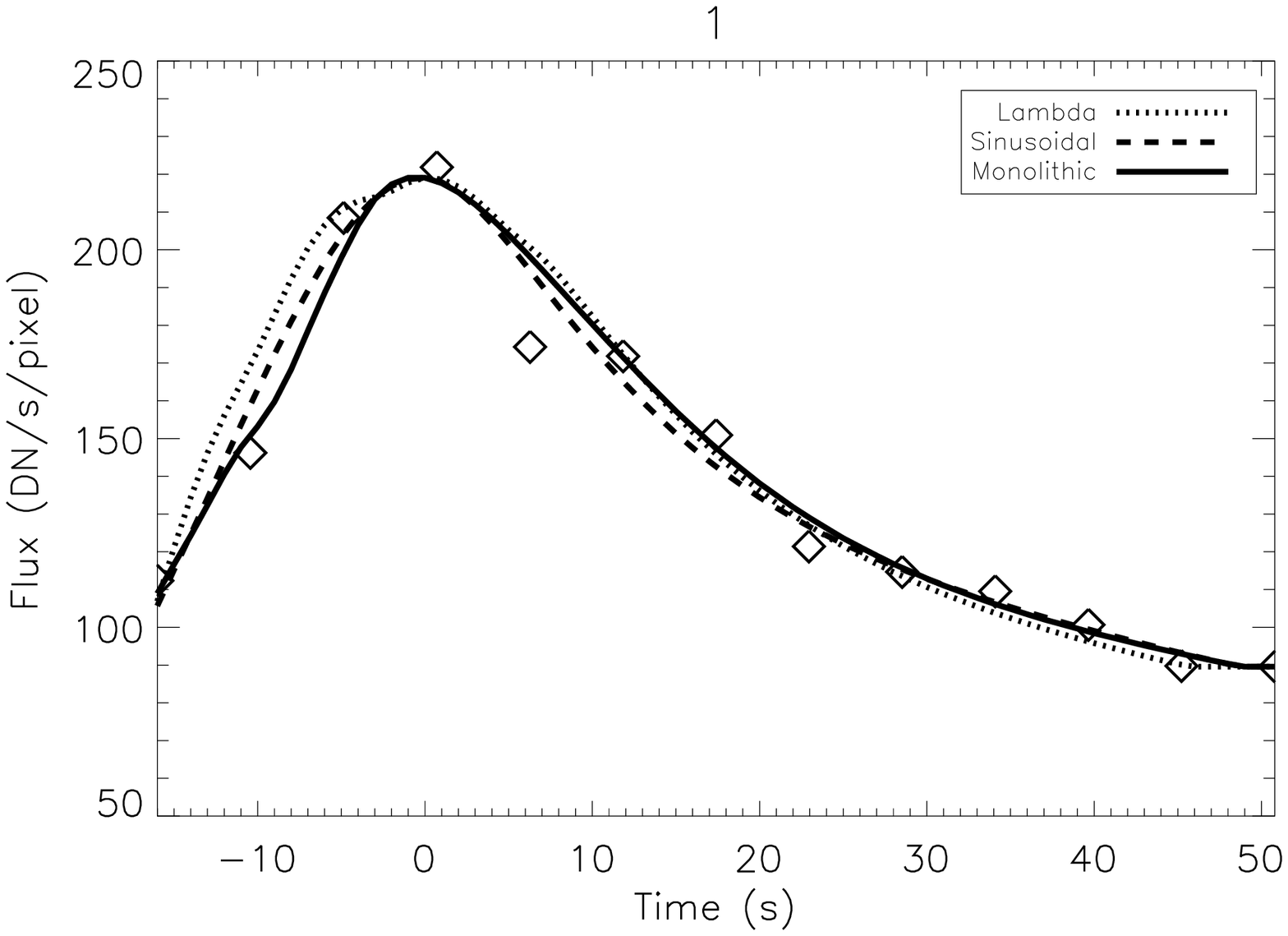}\includegraphics[width=0.5\linewidth]{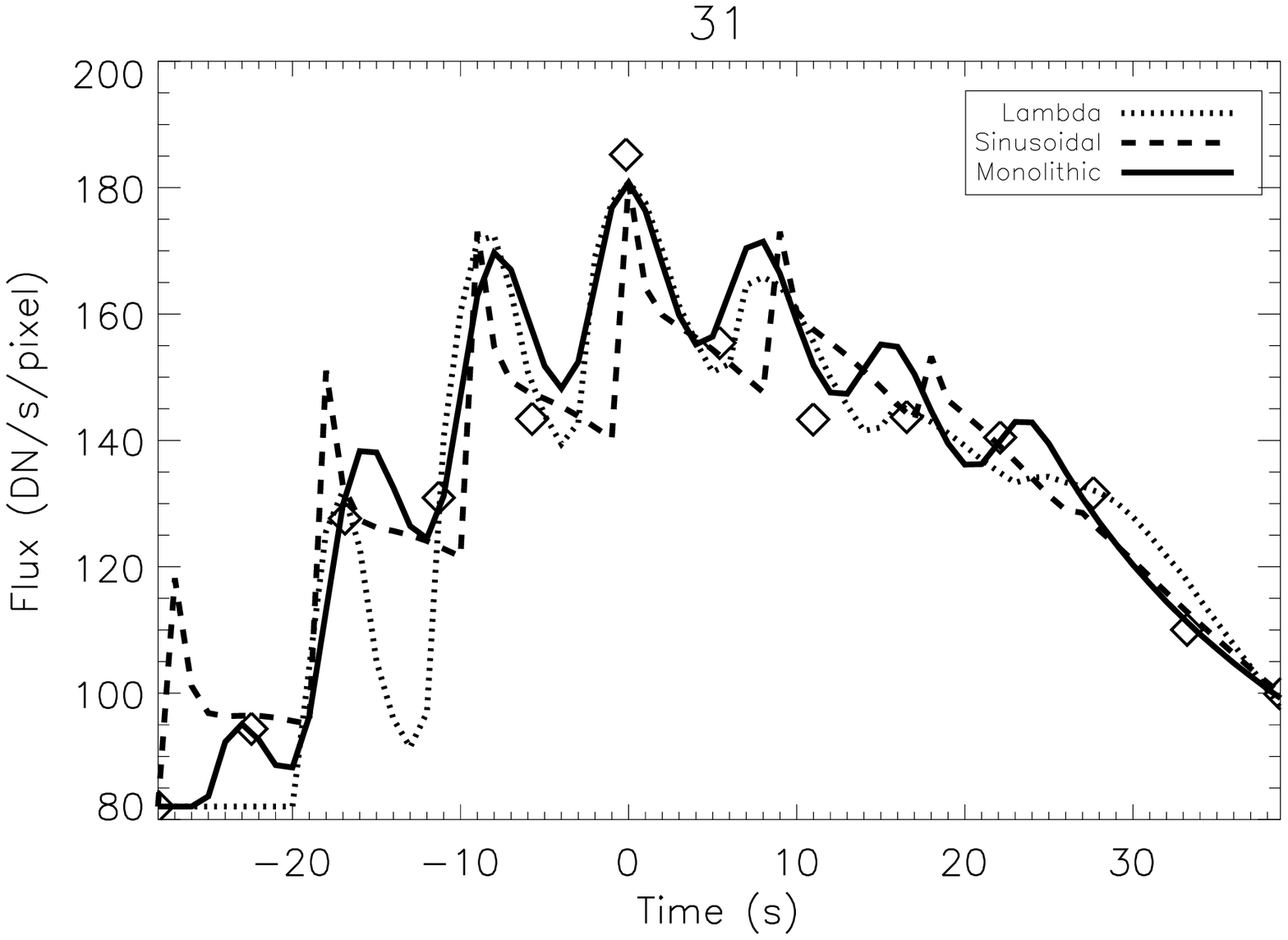}
\caption{Example fit of the flux for ARTB numbers 1 and 31 as listed in Table~\ref{hic_artb_tab}. The dotted line is the fit from the lambda envelope, the dashed for the sinusoidal envelope and black for the monolithic envelope. The observational flux is denoted with diamonds. As was typical of the fits for Hi-C data, all three models visibly fit the observed light curve quite well. The quality of the fits is also showcased by the low $\chi^2$ values ($<$30 DN) and integral ratios ($<$ 3\% difference) in Table~\ref{hic_rslts_temp_tab}. The left fit is a typical fit, the right was chosen to illustrate the potential dynamics that our model could capture.}
\label{hic_fit_examp}
\end{figure*}

\begin{figure*}[ht]
\includegraphics[width=0.5\linewidth]{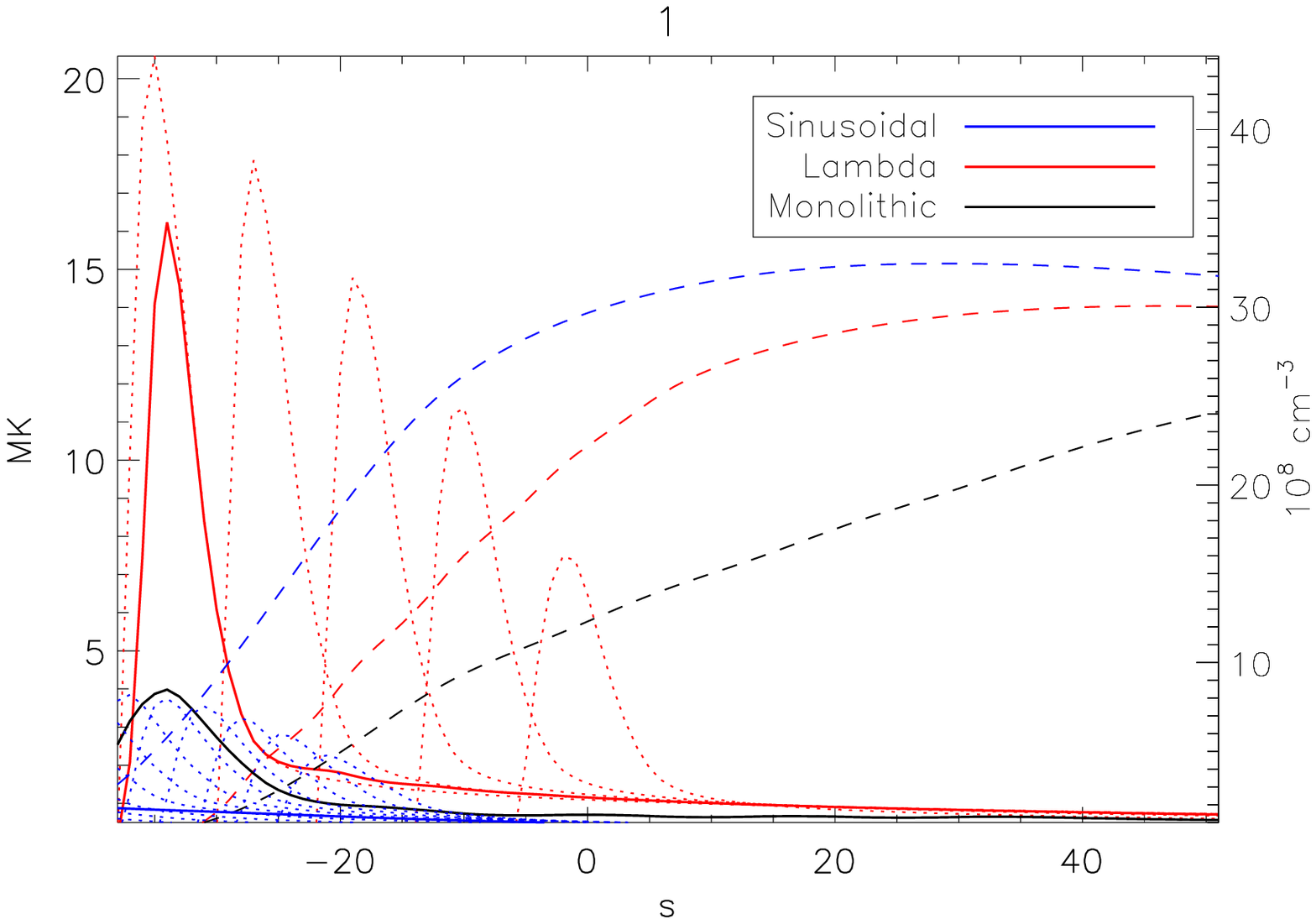}\includegraphics[width=0.5\linewidth]{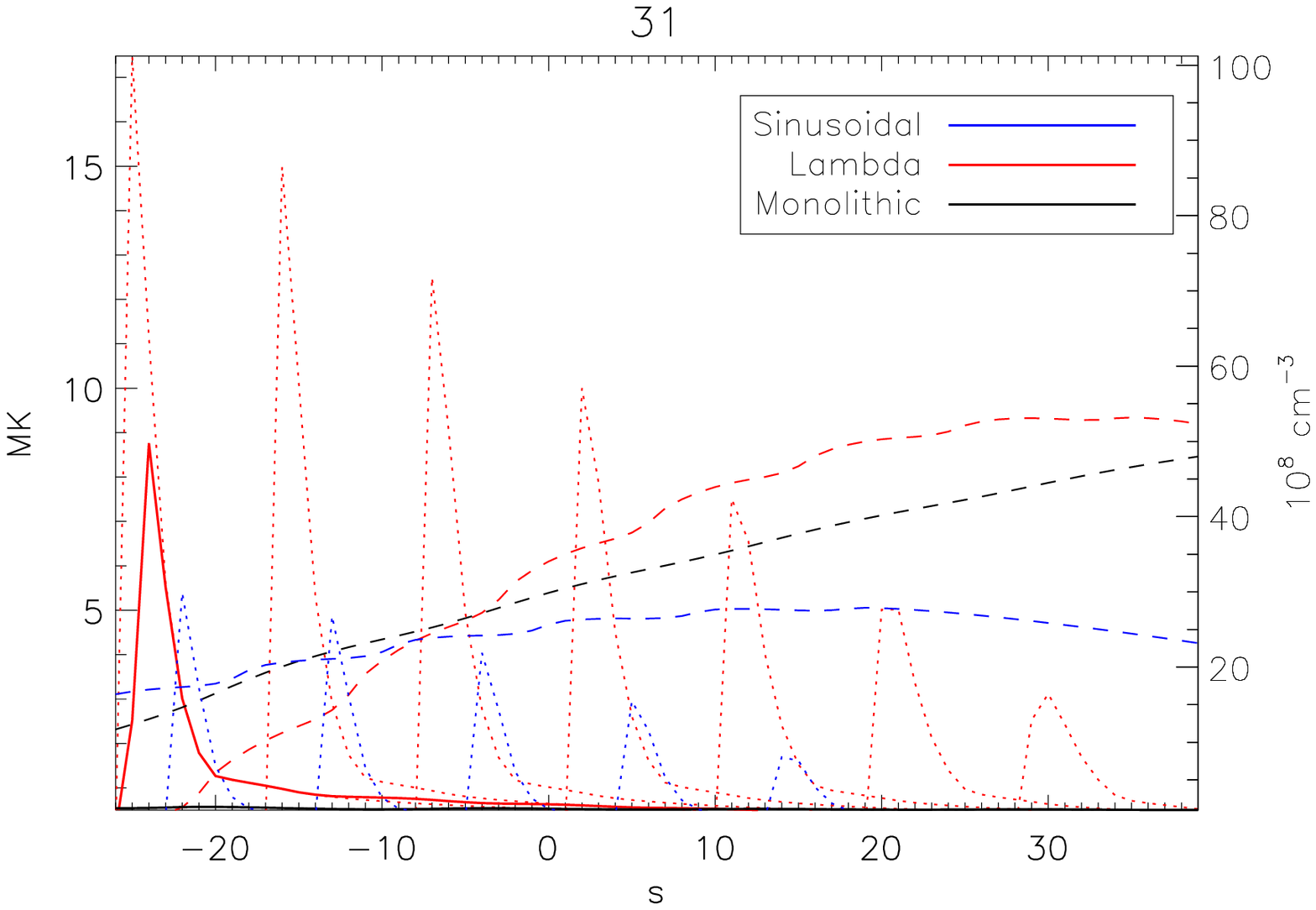}
\caption{Example temperature and density results from the model. These particular results are for ARTB numbers 1 and 31 as listed in Table~\ref{hic_artb_tab}. The dotted curves represent the temperature evolution for each individual strand, while the solid lines are the emission measure weighted temperature. The dashed lines show the evolution of the density for each model. Blue curves are for the sinusoidal envelope, red for the lambda envelope, and black for the monolithic strand. The comparison between the peak temperatures for each envelope is not typical, though the short duration of the elevated temperatures is fairly typical (see Table~\ref{hic_rslts_temp_tab}). Note that for ARTB 31, the sinusoidal envelope result does not get above 1 MK during the observation. These are the same ARTBs and timings depicted in Figure~\ref{hic_fit_examp}.}
\label{hic_temps}
\end{figure*}

\begin{table*}\scriptsize\centering
\caption{ Model Parameters for the results shown in Figures~\ref{hic_fit_examp} and \ref{hic_temps}.}
\begin{tabular}{c|c|c|c|c|c|c}
\hline\hline
 Envelope & Heat Input     & Heat Pulse & Event      & Number of & Half-Length & Strand          \\
  		& (10$^{23}$ ergs) & Width (s)     & Delay (s) & Heatings   & Result (Mm)& Radius (km) \\
 \hline
 ARTB number 1\\
 \hline
Lambda & 3.37 & 11 & 8 & 26 & 1.53 & 107\\
Sinusoidal & 3.44 & 20 & 3 & 15 & 1.37 & 66\\
Monolithic & 3.39 & 8 & 8 & 25 & 1.69 & 257\\
 \hline
  ARTB number 31\\
 \hline
 Lambda & 0.83 & 11 & 9 & 25 & 0.33 & 79\\
Sinusoidal & 1.90 & 5 & 9 & 12 & 0.38 & 129\\
Monolithic & 1.44 & 6 & 8 & 1 & 0.93 & 133\\
\hline
\end{tabular}
\label{hic_exampfits_params}
\end{table*}

\begin{table*}\scriptsize\centering
\caption{Median of the best fit parameters (and median absolute deviation) used for the model for each envelope after running the fitting algorithm on the detections shown in Table~\ref{hic_artb_tab}. The average observed half-length was 1.18$\pm$0.12 Mm. For the lambda and sinusoidal case, the number of heatings represents the number of strands used. The monolithic envelope uses a single strand heated multiple times, with the heating events dictated by the lambda shaped envelope. The aspect ratio is calculated as the resultant strand half length divided by its radius.}
\begin{tabular}{c|c|c|c|c|c|c|c}
\hline\hline
 Envelope & Peak Heating 		  & Heat Pulse & Event      & Number of & Half-Length & Strand & Aspect \\
  		& (ergs cm$^{-3}$s$^{-1}$) & Width (s)    & Delay (s) & Heatings    & Result (Mm) & Radius (km) & Ratio \\
\hline
 Lambda & 0.166$\pm$0.108 & 18$\pm$7 &  6$\pm$2 & 16$\pm$7 & 1.68$\pm$0.89 &  93$\pm$43 &  17.5$\pm$12.2 \\
 Sinusoidal & 0.343$\pm$0.153 & 15$\pm$ 8 &  7$\pm$1 & 17$\pm$7 & 1.85$\pm$0.69 &  52$\pm$37 &  27.5$\pm$19.3 \\
 Monolithic & 0.123$\pm$0.046 &  4$\pm$ 2 &  7$\pm$1 & 17$\pm$8 & 1.73$\pm$0.63 & 107$\pm$62 &  17.6$\pm$11.0 \\
\hline
\end{tabular}
\label{hic_rslts_tab}
\end{table*}

\begin{table*}[ht]\scriptsize\centering
\caption{Median results (and median absolute deviation) from the model. The heat input is the heating function integrated across the entire loop and lifetime of the heating event. The temperature refers to the emission measure weighted temperature. The FWHM is the full-width at half maximum of the temperature profile, and the time above refers to the time the temperature is above 5 MK (both interpolated to 0.01 second resolution). The density is the peak average density from all strands, which does not occur during the temperature peak.}
\begin{tabular}{c|c|c|c|c|c|c|c}
\hline\hline
 Envelope & Integral & $\chi^2$ & Heat Input & Temperature & FWHM & Time Above & Peak Density \\
  & Ratio & (DN) & (10$^{23}$ ergs) & Peak (MK) & (s) & 5 MK (s) & (10$^{8}$ cm$^{-3}$) \\
\hline
 Lambda & 1.007$\pm$0.020 & 24.3$\pm$14.0 &   6.4$\pm$3.3 &  8.3$\pm$3.6 &  7.8$\pm$3.1 &  5.0$\pm$4.5 & 14.2$\pm$ 7.3\\
 Sinusoidal & 1.015$\pm$0.016 & 24.2$\pm$15.7 &   5.1$\pm$3.2 &  3.3$\pm$0.6 & 16.1$\pm$10.5 &  - & 25.6$\pm$13.2\\
 Monolithic & 1.009$\pm$0.009 & 15.2$\pm$ 9.7 &  3.5$\pm$2.2 & 10.8$\pm$3.3 &  8.1$\pm$3.1 &  9.7$\pm$4.3 & 46.6$\pm$20.0\\
\hline
\end{tabular}
\label{hic_rslts_temp_tab}
\end{table*}

It is worthwhile to note the importance of initial conditions used in the EBTEL model. The rate of background heating and strand length in EBTEL determines the initial temperature and density of the loops. When the initial temperature and density are low, the strand temperature can become exceedingly hot, (over 15 MK), but quickly drops down to a more stable temperature before the density is sufficient for any observable signature. In the case of a multi-stranded loop and emission measure weighted temperatures, the staggering of strands will mitigate this effect for all but the first strand, as the short lived high temperatures will be normalized by the densities of the already heated strands. Some of these effects can be seen in Figure~\ref{hic_temps}. Altering the background heating rate to 5\% of the peak of the individual heating functions and re-fitting the best fit results did not change the resultant flux. Above 5\% of the peak of the individual heating events only a few of the results were affected. For these reasons and since we are looking for notable brightenings above the background, we have constrained the background heating to be less than 5\% of the peak heating rate. Additionally, higher background heating rates than used here would create steady loops hotter than observed in \citet[][]{Winebargeretal2013}.

The width of the individual heating pulses in the best fits was smallest for the monolithic envelope, though this number is slightly misleading, as the monolithic loop received a median of 17 heating events, significantly increasing the amount of time the loop experienced heating. This heating pulse width is smaller than the event delay, suggesting that the model did not favor a steady heating in these strands, as there was generally time for the loop to cool slightly before being heated again (see Figure~\ref{hic_fit_examp}). The heat input in Table~\ref{hic_rslts_temp_tab} was calculated by integrating the heating function of all strands in time across the entire volume of the loop, assuming the strands to be cylinders with the lengths and radii returned by the model.

The modeling algorithm was quite adept at fitting all of the detections as shown by the low $\chi^2$, and integral ratio values near unity in Table~\ref{hic_rslts_temp_tab} and Figure~\ref{hic_fit_examp}. The right-hand plot of Figure~\ref{hic_fit_examp} shows the dynamics possible with this model, even though the parameters of this fit might be unnecessarily dynamic due to possible signal oscillations from an uncorrected noise source (such as photon counting). A few of the fits exhibit this type of oscillatory behavior, but not the majority (a more typical example can be found in left panel of Figure~\ref{hic_fit_examp}). 

The multi-stranded sinusoidal envelope resulted in a median peak temperature of over 3.3 MK, while the lambda envelope resulted in median peak temperatures of 8.3 MK and 10.8 MK for the multi- and single-stranded cases respectively. While these results corroborate the existence of hot quiescent active region plasma (above 5 MK), the sinusoidal results are lower than might be expected (see right panel of Figure~\ref{hic_temps}). Additionally, it was shown in testing \citepalias[][]{MTL1} that EBTEL tends to slightly overestimate temperatures when the time step is too large, but our time step size (1s) was empirically chosen to minimize this effect. It is important to note that for all results, the median duration of high-temperature ($>$5 MK) plasma is less than 10s (below the temporal resolution of current instruments), occurs when the strand density is low (Figure~\ref{hic_temps}), and extends over a region smaller than the spatial resolution of most instruments, all of which complicate the detection of the hot component suggested by the model. The strand peak densities are as expected, and occur much later than the temperature peak, minimizing the visibility of the temperature peak in the observed light curves.

By adjusting the peak flux returned by the model to that of the observation (and assuming cylindrical strands that bisect the pixel) we have obtained the crude estimates of the radius of the strands shown in Table~\ref{hic_rslts_tab}. The results suggest a strand size of order 100 kilometers. This crude estimate is less than or comparable to the resolving power of Hi-C ($\approx$150km), and due to the assumptions involved in its calculation should only be used as an order of magnitude estimate. That notwithstanding, these results are consistent with the findings of \citet{CirtainHiC2013, Shimizuetal1992, Winebargeretal2013, Brooksetal2013}, and \citetalias{MTL1} (among others).

\section{Discussion}\label{hic_discussion}

The often quoted ARTB occurrence rate is 1-40 events per hour per active region \citep[][]{BerghmansMcKenzieClette,Shimizuetal1992}, which would predict no more than 2 detections during the observing period of Hi-C, significantly fewer than detected here. Hi-C observed a fairly typical active region, and extrapolating our data (57 detections in an active region in $\approx$5 mins) suggests the possibility of many hundreds per hour. Given the rudimentary nature and small sample size of this observation, this number cannot be taken too literally. It does suggest, though, that many of the events occur at spatial and temporal scales much lower than previous and current EUV detectors can detect, which may have inadvertently biased previous estimates. This was also noted by \citet{Testaetal2013}.

A possible reason for this discrepancy can be seen in Figure~\ref{hic_detect_map}; the detections are not limited to the core of an active region, such as is traditionally studied \citep[][\citetalias{MTL1}]{Shimizuetal1992,BerghmansMcKenzieClette,Seaton2001}. This may bias our detected numbers, but not enough to eliminate the possibility of a higher microflare occurrence rate than previously thought. This may also bias the sample of fitted data, as brightenings outside of active region cores (moss, plage, etc...) could be caused by slightly different mechanisms. 

The resolution of Hi-C very likely plays a significant factor in the number of detected events, as previous studies used significantly less sensitive instruments, both in temporal and spatial resolution. \citet{Shimizu1995} used SXT data which has a pixel size of $\approx$2.5'', and a sensitivity to significantly higher temperatures \citep[$\approx$ 10 MK,][]{Takeda2011}. To test the effect of spatial resolution on ARTB detection, we rebinned the Hi-C data to AIA (0.6" pixel$^{-1}$) and XRT (1.0'' pixel$^{-1}$) pixel scales. For the binned data to be detected by our algorithm, the brightening would have to span at least 1000 Hi-C pixels at XRT resolution, and over 360 Hi-C pixels at AIA resolution (since the algorithm requires a 10 pixel enhancement).  We only detected two coherent brightenings with the AIA scaling, and 0 with the XRT scaling, which is consistent with the expected occurrence rate of 1-40 per hour. This strongly suggests that the rate of these events is significantly higher (possibly 10 times larger) than may have been previously thought: but due to the small number of high resolution observations, it may be premature to revise the projected rate of microflare occurrence with just the Hi-C results. 

Microflares are generally quoted as having a thermal energy content between 10$^{25}$-10$^{29}$ ergs \citep[][]{Shimizuetal1992,BerghmansMcKenzieClette}, but the events studied here appear to be at least an order of magnitude less energetic than those often studied (median values of order 5$\times$10$^{23}$ ergs for all envelopes - Table~\ref{hic_rslts_temp_tab}). Given the improved resolution, it is not surprising that we would detect brightenings with much lower energy contents than previously observed. This also helps to explain the more frequent detection than predicted. 

When the current method of detection and modeling was performed on XRT data \citepalias{MTL1} understandably different results were returned; the detection frequency was lower and the temperatures were higher for XRT. The peak emission measure weighted temperatures found in the XRT results were of order 15-20 MK, which is larger than those found here, and likely attributable to the higher temperature plasmas to which XRT is sensitive. The low densities suggested here would also be difficult to detect with XRT, as densities of order 10$^8$ cm$^{-3}$ require exposure times of order minutes \citep[as noted in][]{Narukage2011}. The visible quality of the fits when using XRT was also less robust than the results here (median $\chi^2$ values 30\% lower and median integral ratios 150\% closer to unity with Hi-C), which is likely due to a combination of factors including the broader passbands of XRT causing more background noise, and the inherent detection bias for larger and more complicated events in XRT.

The previous work with XRT was unable to clearly distinguish between a single strand heated multiple times and multiple strands each experiencing a single heating event. While there is still some ambiguity with the Hi-C results, it seems apparent from the $\chi^2$ and integral ratios shown in Table~\ref{hic_rslts_temp_tab} that at these spatial and temporal scales, a few to tens of strands are being heated, but each is heated multiple times. Additionally (but subjectively), the visual quality of the fits were most often best when using the monolithic loop, though this is statistically unsupported. This preference for fewer strands can potentially be used to improve the multi-thermal model by establishing a firmer spatial and temporal scale to allow the study of multi-stranded multiply heated loops observed by instruments with lower resolutions than Hi-C. The monolithic loops will be above 5 MK for a longer period of time, since the multiple heating events keep their temperatures high. They also tend to be the most dense of the three envelopes, which promotes the visibility of these structures. In fact, a single strand was preferred by the model even when multiple strands were allowed on two occasions: one case using the lambda envelope and one case using the sinusoidal envelope.

A useful comparison is to the work of \citet{Winebargeretal2013}, who studied different loop brightenings in Hi-C. Combining the Hi-C data with AIA, they found the loops to be cool ($\approx$0.25 MK) and dense (10$^{10}$ cm$^{-3}$), suggesting there was not hot plasma within the observed loops. The cool temperatures were inferred from the fact that the brightenings appeared in both hot and cool channels of AIA simultaneously, and were supported by a DEM analysis. Due to the very short duration of the high temperature phase suggested in the present work, it would be difficult for the microflares analyzed here to show up in AIA, given the lower cadence and low strand filling factor. These discrepancies could result simply from analyzing a different type of event, as the loops studied herein are not necessarily as easily identified as loops given the automated detection algorithm employed. The objectives of the present work differ from those of \citet{Winebargeretal2013} (as well as those of \citet{Brooksetal2013,Brooks2012}), as reflected in the methodologies involved: \citet{Winebargeretal2013} and \citet{Brooksetal2013} were specifically investigating the nature of isolatable loops, and so sought out features that could unambiguously be identified as such for their analyses; whereas the present work focuses on transient brightenings regardless of their shape, and thus utilizes an objective detection algorithm that triggers only on changes in the brightness.

Additionally, we also note the correspondence between the energy input and the observed radiation. It is common to use the observed radiation as a proxy for the total energy of the microflare \citep[i.e.,][]{Hudson1991}, and we can now compare the models energy input into the strand to the various radiative measurements. In Figure~\ref{hic_nrg_obs} we compare the observed radiation flux, input energy, and coronal radiative losses. The radiation flux is calculated by converting the detected light curve flux into energy by assuming all observed light is 193\AA\ and converting from DN s$^{-1}$ pixels$^{-1}$ to ergs s$^{-1}$ pixels$^{-1}$ using the instrument response function of Hi-C. The losses are calculated from the EBTEL results for the best fit of each observation. These results suggest the radiative losses scale with the input energy using a power law with an exponent between 0.65 and 0.75, though this relation does not carry over to the observed radiation. For a given input energy, the observed radiation can be found over as many as two decades in energy, illustrating the difficulty of using observed flux as a direct proxy for energy input in flaring events. This discrepancy between the observed radiative losses is amplified when using a narrow band instruments such as Hi-C.

\begin{figure*}[ht]
\includegraphics[width=1.0\linewidth]{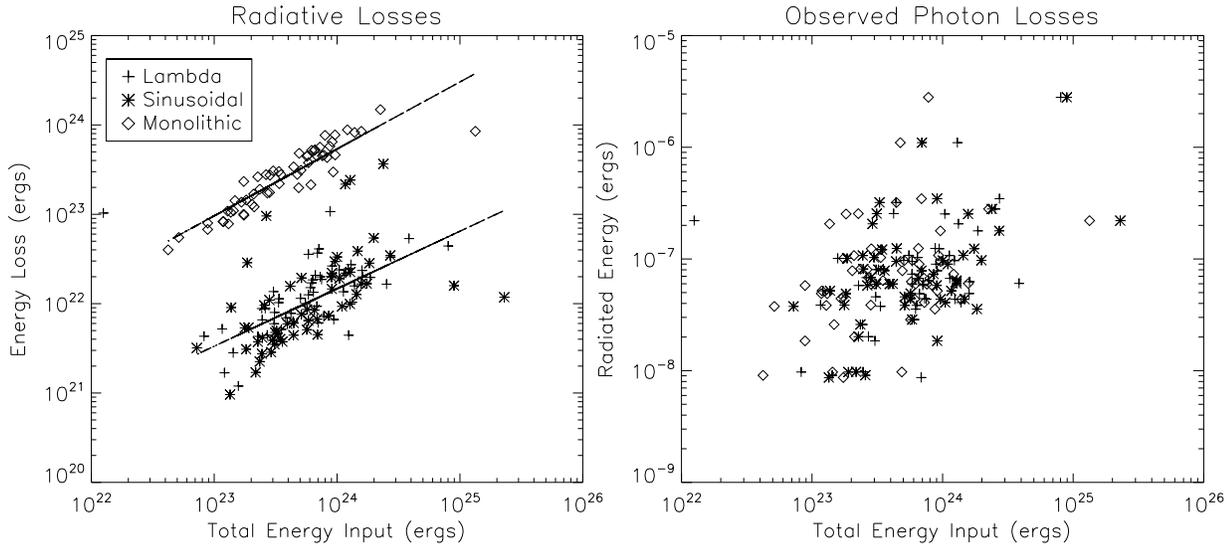}
\caption{Coronal radiated energy distributions from the modeling and observations. The horizontal axis in all plots is the total integrated energy input of the model, less background flux. The lines in the left plot represent a best fit power law function between the integrated radiative loss function and the total energy input for the single and multi-stranded cases, suggesting a power law ratio between them with an exponent between 0.65-0.75. The left compares the total energy to the integrated coronal radiative loss function. The right shows the integrated energy incident on the detector, with background subtracted, and assuming a wavelength of 193\AA~for the incident photons. The lack of any distinctive trend in this plot illustrates the difficulty in using the observed radiation as a proxy for the energy input.}
\label{hic_nrg_obs}
\end{figure*}

\section{Conclusion}\label{hic_conc}

We have detected and modeled 57 brightenings observed with the Hi-C instrument as strands experiencing different forms of heating, and found them to be quickly and consistently heated to temperatures above 5 MK when the density is still low ($10^8$ cm$^{-3}$), generally cooling to temperatures below 5 MK very quickly, before there is significant density for direct detection. In general, the densities peaked at 10-40$\times10^{8}$ cm$^{-3}$ after the temperature had returned below 5 MK. The size of the strands is also sub-resolution for most instruments, which makes detection difficult without Hi-C, whose pixel size is significantly closer to the size of the strands, but still not quite small enough to be resolved.

The results from this study provide an improvement from those of \citetalias{MTL1}, as the monolithic envelope may be marginally preferred by the model at Hi-C scales. The model was able to match the observed data more easily (compared to the results from \citetalias{MTL1}) with very reasonable parameters (strand widths, densities, temperatures, radii, etc...). The widths of these strands should be near the resolution of the next generation of imagers (including Hi-C, EUI on Solar Orbiter and the proposed XIT on Solar-C), and thus the larger strands could be resolved on a consistent basis.

The number of detections of coherent brightenings was significantly higher than predicted by previous estimates, which is likely due to the unprecedented resolution of Hi-C. The length of time of the Hi-C observations and sample size is too small to effectively update the predicted frequency of these events for future observations, though we recommend that more observations at similarly high temporal and spatial resolution be made to explore more thoroughly the frequency of their occurrence, as well as the distributions of size, duration, and energy budget. With more results, we could better develop the distribution between the frequency of events and energy input of microflares. Better understanding of this distribution would improve the discussion of whether nanoflares can heat the corona.

In all, these results strongly promote the scientific utility and benefit of higher resolution coronal imagers, such as Hi-C.

\acknowledgments
We acknowledge the High resolution Coronal Imager instrument team for making the flight data publicly available. We especially thank Sabrina Savage for her help with acquiring and calibrating the data. We also thank Hugh Hudson for helpful discussions about microflares. We also thank Dana Longcope and the anonymous referee for their thoughtful input which improved the manuscript. This work was partially supported by NASA under contract NNM07AB07C with the Smithsonian Astrophysical Observatory.

\bibliographystyle{apj}       
\bibliography{kobelski}

\clearpage
\end{document}